# Electro-Chemo-Mechanical Modeling of Multiscale Active Materials for Next-Generation Energy Storage: Opportunities and Challenges

**Dibakar Datta**
Department of Mechanical and Industrial Engineering
New Jersey Institute of Technology (NJIT), Newark, NJ 07103, USA
Email: dibakar.datta@njit.edu

**Abstract:**

The recent geopolitical crisis resulted in a gas price surge. Although lithium-ion batteries represent the best available rechargeable battery technology, a significant energy and power density gap exists between LIBs and petrol/gasoline. The battery electrodes comprise a mixture of active materials particles, conductive carbon, and binder additives deposited onto a current collector. Although this basic design has persisted for decades, the active material particle's *desired size scale* is debated. Traditionally, microparticles (size range $\geq 1 \, \mu m$) have been used in batteries. Advances in nanotechnology have spurred interest in deploying nanoparticles (size range 1-100 nm) as active materials. However, despite many efforts in *nano*, industries still primarily use 'old' microparticles. Most importantly, the battery industry is unlikely to replace microstructures with nanometer-sized analogs. This poses an important question: *Is there a place for nanostructure in battery design due to irreplaceable microstructure?* The way forward lies in multiscale active materials, microscale structures with built-in nanoscale features, such as microparticles assembled from nanoscale building blocks or patterned with *engineered* or *natural* nanopores. Although experimental strides have been made in developing such materials, computational progress in this domain remains limited and, in some cases, negligible. However, the fields hold immense computational potential, presenting a multitude of opportunities. This perspective highlights the existing gaps in modeling multiscale active materials and delineates various open challenges in the realm of electro-chemo-mechanical modeling. By doing so, it aims to inspire computational research within this field and promote synergistic collaborative efforts between computational and experimental researchers.

## ToC Graphic

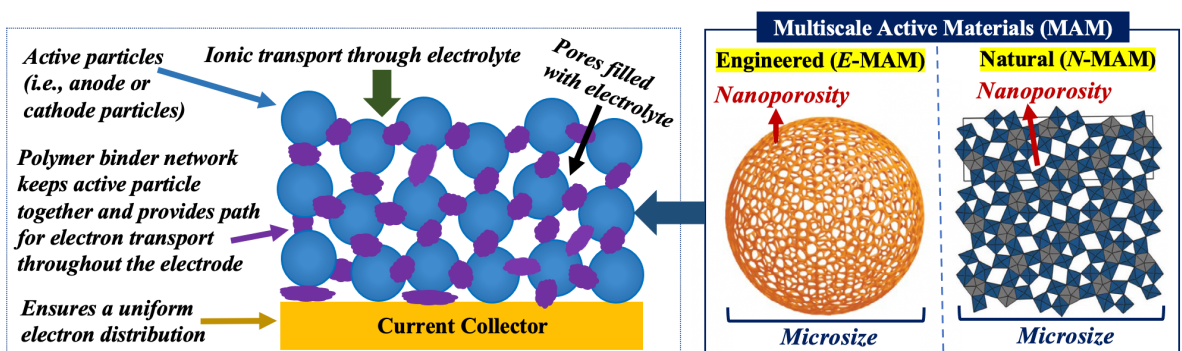

**Multiscale Active Materials (MAM) as electrode for next-generation energy storage system**

## 1. INTRODUCTION

Rechargeable batteries have become an integral part of our daily lives, finding diverse applications in portable electronics, electric vehicles, grid energy storage, and renewable energy systems[1]. To drive the advancement of modern society, the demand for efficient, affordable, and safe energy storage solutions is paramount[2]. Presently, rechargeable lithium-ion batteries (LIBs) stand as the dominant technology in energy storage, yet a considerable disparity persists in both energy and power density when compared to conventional fuels like petrol[3]. Thus, the development of novel high-rate electrode materials capable of rapid charge storage within minutes, rather than hours, is imperative to enhance power output and reduce charging time in LIBs[4, 5]. These materials hold the potential to address challenges tied to electric vehicle adoption, grid-scale energy storage, and the creation of high-power devices[6].

The energy and power density of LIBs are intricately linked to the composition and particle size of active electrode materials, which in turn influences the electrode fabrication process[7]. Hence, the deliberate selection of active electrode materials with specific particle sizes stands as a pivotal considerable in battery research. This issue has been extensively explored by Jain et al.[7], who analyzed the concept of nanostructured and microstructured electrodes. Nanostructured electrodes encompass active material particles within the 1-100 nm size range, while microstructured electrodes employ particles of micrometer size ($\geq 1$ $\mu m$). Presently, industrial applications predominantly employ microstructure-based electrodes[7]. Over the past decades, significant research has been dedicated to utilizing nanomaterials for energy storage[8, 9]. Nanostructured electrodes offer distinct advantages such as heightened high-rate performance, increased power density, enhanced lithium solubility and capacity, diminished memory effects, as well as improved fracture resilience and fatigue resistance[7].

Despite the widespread implementation of strategies like nanostructuring to bolster high-rate performance in materials like LTO ($Li_4Ti_5O_{12}$), such approaches come with drawbacks including elevated cost, instability, reduced volumetric energy density, low initial Coulombic efficiency, limited mass loading, high manufacturing complexity associated with nanoparticles[7]. As a result, industries remain cautious about direct replacement of nanostructured electrodes with microstructured ones. However, microstructured electrodes have their own limitations like lower gravimetric capacity and reduced fracture resistance[7].

Jain et al.[7] have proposed a forward-looking strategy that advocates for the utilization of Multiscale Active Materials (MAM), combining the features of both nanostructures and microstructures. These MAM can either be engineered or natural. Engineered MAMs (E-MAMs) might encompass microparticles containing purpose-engineered nanopores or composites of nanoparticles and 2D materials. Natural MAMs (N-MAMs) refer to naturally occurring oxide materials such as Niobium Tungsten Oxides (NTO) and Molybdenum Vanadium Oxides (MVO). Recent studies have demonstrated the achievement of high rates in micrometer-sized particles of complex oxides of niobium (T-$Nb_2O_5$), ternary Nb/W oxides ($Nb_{16}W_5O_{55}$ and $Nb_{18}W_{16}O_{93}$), and ternary Ti/Nb oxides ($TiNb_{24}O_{62}$ and $TiNb_2O_7$)[10].

While experimental endeavors[11-13] have showcased the remarkable potential of MAMs for next-generation energy storage, progress on the modeling front has been relatively limited[14]. Computational challenges have hindered the advancement of MAM modeling. However, there exists an urgent need for modeling progress to complement experimental efforts. For instance, Lakhnot et al.[14] recently compared different intercalation compounds and highlighted the potential of layered oxides and MAM open-tunneled oxides for multivalent ion insertion. Yet, their analysis focused on existing open-tunneled oxide materials, leaving room for the exploration of novel compounds. This quest is akin to finding a needle in a haystack, given the vast material possibilities. Traditional trial-and-error experimental approaches are unlikely to yield optimal solutions, underscoring the urgency of computational advances in the field of MAM.

This perspective provides a concise overview of the limited strides made in MAM modeling, subsequently emphasizing the immense opportunities and concurrent challenges within this emerging domain. The structure of the paper is as follows: Section 2 succinctly outlines the pros and cons of nanomaterials, while Section 3 delves into diverse MAMs for potential future applications. Section 4 offers a brief snapshot of the current status of MAM modeling. In Section 5, potential computational challenges in modeling MAM are elucidated. Section 6 takes a brief look at the modeling challenges pertaining to MAM. Finally, Section 7 concludes the perspective, with the anticipation that this paper will galvanize the computational community to engage with these challenging problems and foster productive synergistic computational-experimental collaborations.

## 2.  ADVANTAGES AND DISADVANTAGES OF NANOMATERIALS

In a concise analysis, the advantages, and disadvantages of nanomaterials for energy storage can be summarized as follows. A more comprehensive elaboration can be found in the work of Jain et al[7].

### 2.1 Nano- versus Microstructure for Battery Electrodes: What are the Advantages of *Nano*?

**Mechanics:** The implementation of nanostructuring yields enhanced mechanical stability when compared to microstructuring[15, 16]. The nanoparticles' diameter exists within or beneath the typical crack size range, mitigating the occurrence of fractures. Nanoparticles exhibit a more uniform intercalation behavior than microparticles, thereby reducing strain inconsistencies and particle fractures. Consequently, this leads to an improvement in the fracture toughness and fatigue life of batteries[17].

**Kinetics:** Nanostructuring proves favorable for swift charging/discharging, high-rate operations, and high-power density[18, 19]. The ion diffusion time is denoted as $\tau = \lambda^2/D$ (where $\lambda$ signifies the transverse length dependent on particle size, $D$ stands for diffusion coefficient)[15, 18]. The maximum attainable $C$-rate limited by bulk diffusion is given by[20] $C_D = 3600 \, D/\lambda^2$. Therefore, transitioning from microparticles ($\lambda \geq 1 \, \mu m$) to nanoparticles ($\lambda \approx 1 \, \mu m$) reduces $\tau$ and enhances $C_D$.

**Thermodynamics:** Microstructured electrodes entail negligible excess surface energy. Nonuniform intercalation can trigger early phase transformations[21]. In contrast, nanostructured electrodes significantly elevate the free energy of constituent phases, consequently altering the voltage profile[22]. Phase transitions occur rapidly, releasing excess free energy originating from lattice mismatch and high surface area[23].

### 2.2 Nanomaterials Have Many Advantages. But Why the Industry Still Prefers Micromaterials?

**Poor Volumetric Energy Density:** While most academic studies emphasize gravimetric energy density (Wh/kg), practical energy storage in confined spaces necessitates maximizing energy within limited volume, emphasizing volumetric energy density (Wh/L)[24]. Nanoparticle-based electrodes suffer from poor packing density due to void spaces between nanoparticles[25].

**Reduced Coulombic Efficiency:** The small size and ultrahigh specific surface area of nanomaterials have dual effects. On one hand, these attributes enhance fracture resistance, fatigue life, and rapid diffusion. Conversely, electrolyte decomposition at nanoparticle surfaces generates substantial solid-electrolyte interphase (SEI) during the initial cycle[26], significantly diminishing coulombic efficiency[27, 28].

**Low Mass Loading and Aggregation:** Industries aspire to achieve high mass loading (20-30 mg/cm$^2$) for electrodes. This goal is unattainable with nanomaterials due to their relatively low tap density[29, 30].

**High Cost and Complexity:** Apart from elevated costs[31], the manufacturing of nanostructured electrodes on a large scale poses challenges[32], leading to the generation of environementally harmful chemical waste[15, 33, 34]. Additionally, nanostructuring necessitates suplementary post-processing[7, 31].

Figure 1 highlights the advantages of micro- and nanoparticles in lithium-ion battery applications.

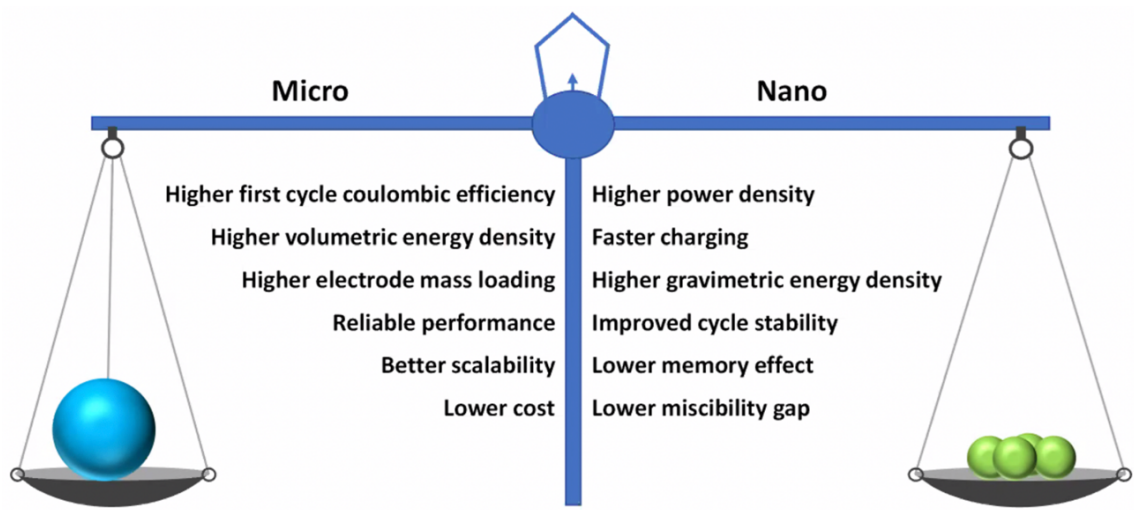

**Figure 1.** Schematic depicting the advantages of micro- and nanoparticles in lithium-ion battery applications. Image Credit: Prof. Nikhil Koratkar (RPI)

## 3.  MULTISCALE ACTIVE MATERIALS (MAM): THE FUTURE

### 3.1 Both Nano and Micro Have Pros and Cons. So What's the *Transformative* Solution?

The realm of both nanoscale and microscale materials carries its own set of advantages and disadvantages. So, what's the groundbreaking solution? The answer lies in Multiscale Active Materials (MAM), a concept that merges microstructures with nanoscale features. By harnessing the appealing attributes of both nanoparticles (NPs) and microparticles (MPs) within a single entity, we can truly leverage the best of both domains. MAM can be classified into two main types:

### 3.1.1 N-MAM : Natural Multiscale Active Materials

Certain microscale materials exhibit inherent nanoscale channels or tunnels. A fascinating example of such materials is complex oxides, a domain relatively less explored by the computational chemo-mechanics community. The likes of Niobium Tungsten Oxides (NTO)[11] structures (Fig. 2a), including $Nb_{16}W_5O_{55}$, $Nb_{18}W_{16}O_{93}$, belong to the Wadsley-Roth type of crystallographic shear structure[10]. Similarly, Molybdenum Vanadium Oxides (MVO)[14] structures (Fig. 2b) come in different polymorphs – orthorhombic ($MoV_2O_8$), trigonal ($MoV_3O_6$), tetragonal ($MoVO_5$), each boasting distinct internal tunnels such as hexagonal, heptagonal, pentagonal, and rectangular configurations.

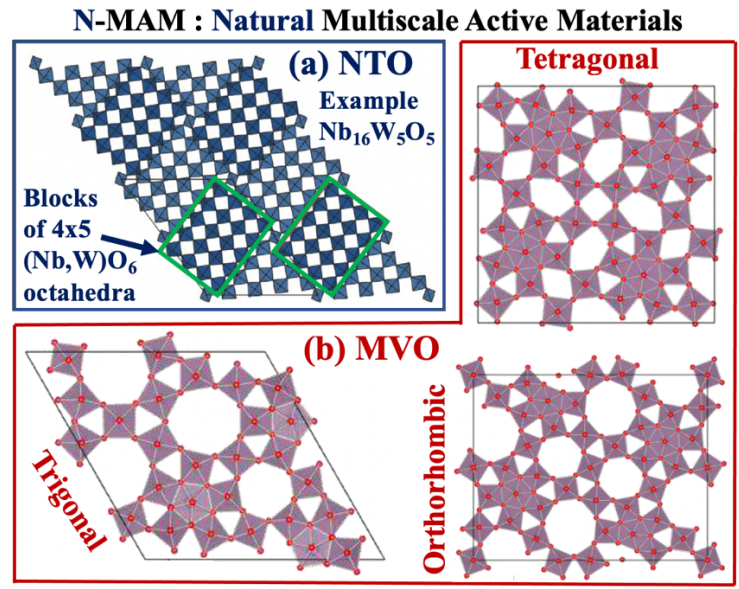

**Figure 2. Examples of *N*-MAM:** Class of **[a]** Niobium Tungsten Oxides (NTO), **[b]** Various Molybdenum Vanadium Oxides (MVO).

### 3.1.2 E-MAM : Engineered Multiscale Active Materials

This category can be divided into two subcategories:

**(i) Engineered Nanoporosity:** Micrometer-sized alloy-based particles can be endowed with nanometer-sized pores through various methods[35, 36].

**(ii) Assembly of Nanoscale Constitutents:** This class encompasses several approaches:

(a) *NP + NP Ensemble:* Various techniques can assemble NPs into larger MPs[37-39] (Fig. 3a).

(b) *NP + 2D Materials:* The realm of 2D materials[40-45] and heterostructures[46, 47] has garnered immense interest in mechanics[48, 49], finding applications in diverse fields including batteries[50-52]. By integrating 2D materials with NPs (Fig. 3b)[53-55], an array of engineered E-MAM structures can be formed, propelling innovation and progress in energy storage.

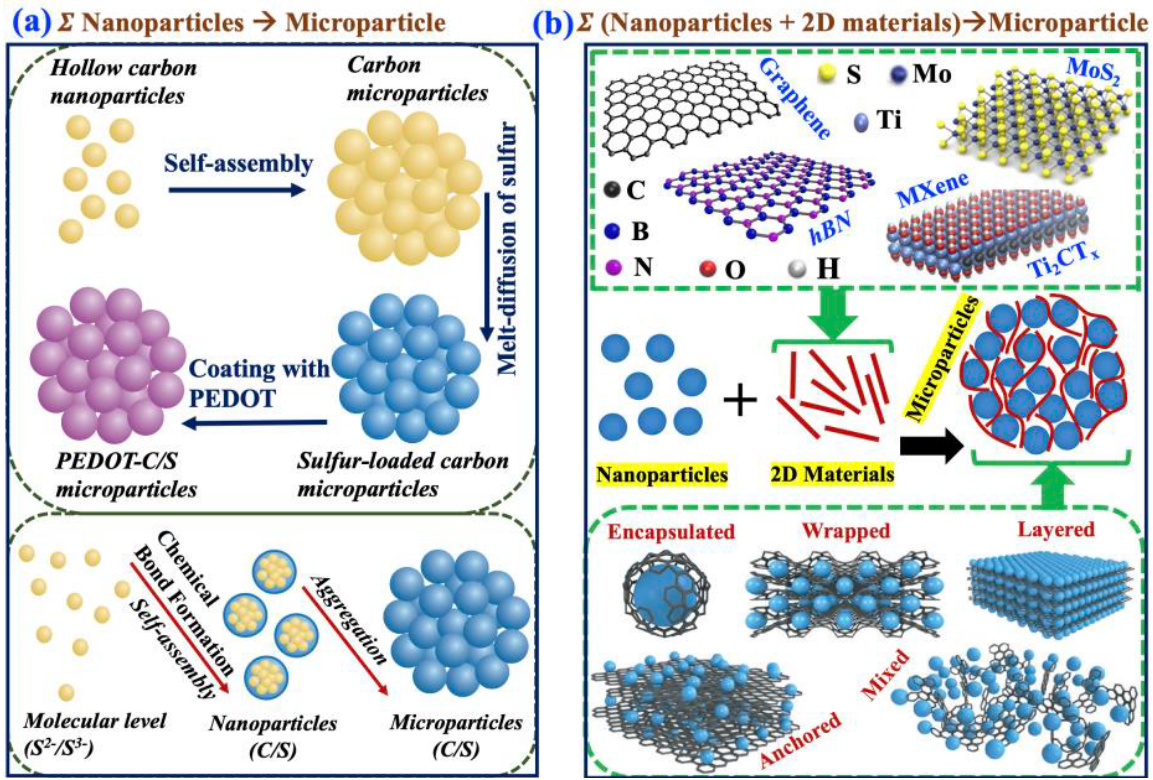

**Figure 3. Examples of *E*-MAM:** **[a]** Microparticles from nanoparticles, **[b]** 2D materials-integrated nanoparticles assembled microparticles.

## 4. MODELING OF MAM: OVERVIEW OF CURRENT STATUS

From an electro-chemo-mechanical perspective, the following four interconnected challenges demand immediate attention:

- **Interfacial Mechanics in E-MAM:** E-MAM comprise composite electrodes constructed from NP-NP or NP-2D material systems (Fig. 3). Consequently, comprehending interfacial phenomena stands as a pivotal factor in effectively designing E-MAM.

- **Mechanics of MAM Electrode-Electrolyte Interface and SEI Formation:** In practical applications, electrodes interact with electrolytes. Addressing the issue of chemically unstable electrodes is imperative, leading to the study of electrode stability within electrolyte environments and the elimination of chemically unstable counterparts. Moreover, for stable electrodes, understanding the formation of SEI and exploring the mechanical characteristics of SEI products carry significant importance.

- **Mechanics of MAM Electrodes During Charge/Discharge:** The process of charge and discharge triggers a range of chemo-mechanical alterations in electrodes, including stress generation, volume fluctuations, and fracturing. Composite electrodes, like E-MAM, experience stress both at the interfaces and within the active materials themselves. Excessive interfacial stress might lead to fractures, delamination, and eventual failure of the active materials. Similarly, N-MAM undergoes various chemo-mechanical phenomena during charge/discharge.

- **Data-Driven Mechanics of MAM electrodes:** Tackling the computational modeling of the aforementioned challenges presents notable complexities. Particularly, performing Density Functional Theory (DFT) simulations proves computationally demanding. Challenges are also apparent in executing Molecular Dynamics (MD) simulations due to the absence of appropriate interatomic potentials. Furthermore, continuum modeling struggles to accurately determine properties at the atomic scale, such as adhesion. Thus, an urgent need arises to develop a machine learning framework capable of efficiently predicting interfacial mechanical properties, the formation of electrode/electrolyte interphases, and the chemo-mechanical attributes of electrodes during the charge/discharge process. This framework would significantly advance the understanding and manipulation of MAM materials.

In summation, addressing these challenges in MAM electro-chemo-mechanical systems not only advances fundamental scientific understanding but also paves the way for innovation that can revolutionize energy storage and related fields.

## 4.1 Progress on Interface Mechanics of MAM

The realm of E-MAM encompasses composite electrodes skillfully crafted from the assembly of NP-NP and NP-2D materials. A fundamental grasp of interface mechanics is pivotal in ensuring the optimal functionality of E-MAM. A central question emerges: how do we judiciously select the most suitable combinations of E-MAM, specifically NP-NP and NP-2D materials assembly? A hypothesis can be posited that the strength of interfaces correlates with interfacial charge transfer, bonding characteristics, and potential gradients. Key parameters for scrutiny encompass interface adhesion, quantified by the work of separation $(W_{sep})$[56-59], between NP-NP and NP-2D materials, and the establishment of a connection between interface adhesion or Work of Separation $(W_{sep})$ and aspects such as interface charge transfer[60, 61], bonding, and potential gradient.

Recent efforts have delved into the examination of 3D/3D and 3D/2D interfaces through atomistic and molecular simulations. Basu et al.[56] explored adhesion (measured by $W_{sep}$) across selected 3D/3D and 3D/2D interfaces. For instance, the combination of amorphous silicon (a-Si) and various substrates (Fig. 4a) was investigated. The $W_{sep}$ value for the a-Si/graphene system was approximately one-fifth of those observed for Cu and Ni substrates, and less than half of the a-Si/a-Si substrate pairing. This suggests that graphene fosters a lubricious interface with the a-Si film. Sharma et al.[58] delved into the correlation between interface strength and charge-transfer (Fig. 4b1) and potential energy gradients (Fig. 4b2). Increased charge transfer was found to signify robust binding.

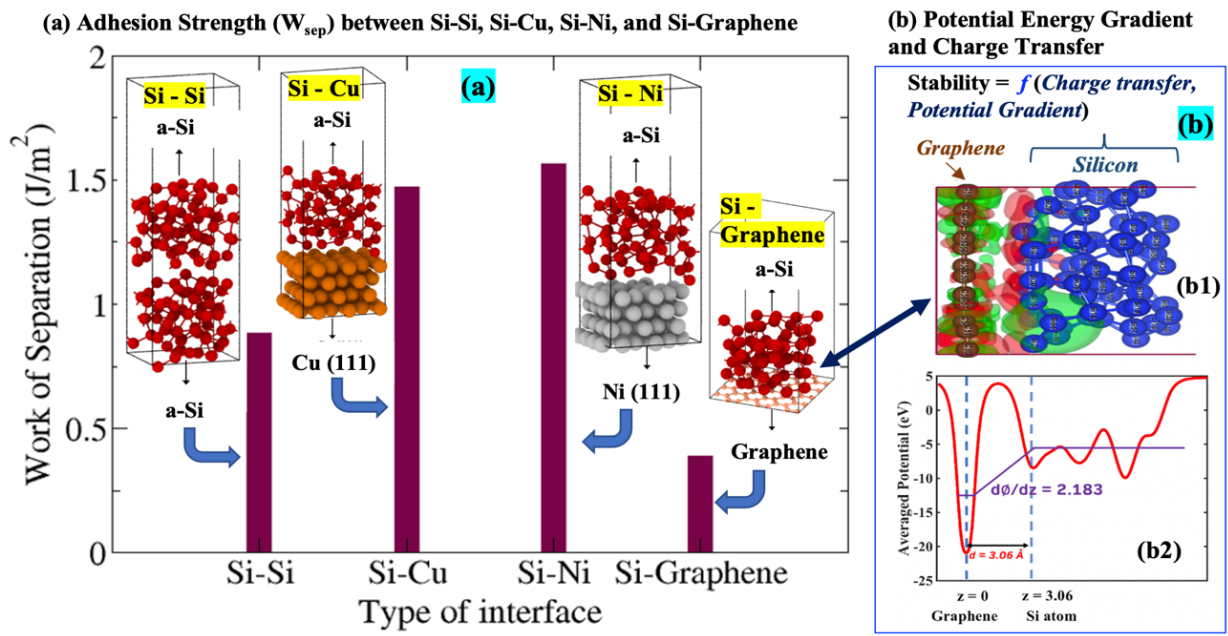

**Figure 4: [a]** Interface adhesion and **[b]** its correlation with (b1) charge transfer and (b2) potential energy gradient. Reproduced with permission from Ref[56, 58].

Sharma et al.[58] recently significantly expanded the understanding of various 3D/2D interfaced systems. Through a comprehensive DFT analysis, they delved into interface strength and bonding mechanisms between crystalline and amorphous selenium (Se) with graphene – a promising duo for energy storage applications. The interface strengths of monoclinic Se (0.43 J/m$^2$) and amorphous Si with graphene (0.41 J/m$^2$) were found to be comparable. While both materials (*c*-Se, *a*-Si) were loosely bonded to graphene through van der Waals (vdW) forces, *a*-Si/graphene exhibited higher interfacial electron exchange, indicative of robust binding.

Sharma et al.[57] further explored the influence of surface-engineered Ti$_3$C$_2$ MXenes on the interface strength of silicon. Different Ti$_3$C$_2$ MXene substrates with surface functional groups (-OH, -OH and -O mixed, and -F) were considered. Results from DFT analyses unveiled that completely hydroxylated Ti$_3$C$_2$ boasted the highest interface strength (0.60 J/m$^2$) with amorphous Si. This strength dwindled with increasing proportions of surface -O and -F groups. In another study, Sharma et al.[59] delved into the implications of a graphene interface on potassiation in a graphene-selenium heterostructure cathode for potassium-ion batteries. A comparison with a graphene-free cathode highlighted the profound structural and electrochemical alterations introduced by a vdW graphene interface to the K$_x$Se cathode.

### 4.2 Progress on Mechanics of MAM Electrode-Electrolyte Interface and SEI Formation

Within the realm of MAM, the interplay between chemical and mechanical attributes is crucial at the electrode-electrolyte interface. Electrodes prone to chemo-mechanical instability can exhibit unfavorable reactions with electrolytes, leading to the production of excessive interphase products and subsequent structural deterioration. For optimal performance, a SEI should possess elasticity and flexibility, allowing adaptation to non-uniform electrochemical behavior and the dynamic changes in active materials.

Therefore, it becomes paramount to explore whether electrodes maintain chemo-mechanical stability when interacting with electrolytes. Those deemed 'unstable' should be discarded. In contrast, for 'stable' electrodes, a comprehensive understanding of SEI formation mechanism and the mechanical properties of the resultant SEI layer is essential. Additionally, the behavior of the electrode/electrolyte interface will inevitably vary based on the nature of electrolyte, whether organic or aqueous.

Kim et al.[62] embarked on MD simulations to dissect the influence of different electrolytes on the structure and evolution of the SEI in LIBs (Fig. 5). Their investigation centered on the formation and expansion of the SEI within a graphite anode, considering ethylene carbonate (EC), dimethyl carbonate (DMC), and mixtures of these electrolytes. The resulting SEI films from EC-rich environments exhibited a significant presence of $Li_2CO_3$ and $Li_2O$, while $LiOCH_3$ dominated the composition of DMC-derived films. Impressively, the computed formation potentials, measuring 1.0 V *vs.* Li/Li+, aligned closely with experimental measurements. Moreover, they evaluated the elastic stiffness of SEI films, revealing their greater rigidity compared to Li metal yet retaining some compliance when compared to graphite anode.

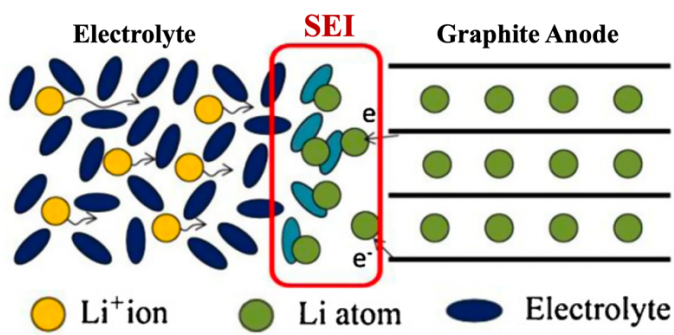

**Figure 5:** SEI (Solid Electrolyte Interphase) formation between graphite anode and EC electrolyte. Reproduced with permission from Ref[62].

However, up to the present moment, there has been a noticeable lack of progress in studying the interface between MAM electrodes and electrolytes. The intricacies inherent in these systems have contributed to this gap in knowledge. Consequently, the study of MAM electrode/electrolyte interactions remains an unexplored frontier brimming with immense potential for breakthrough and discoveries. As researchers continue to delve into this uncharted territory, the insights gained could catalyze transformative advancements in energy storage technology.

### 4.3 Progress on Mechanics of MAM Electrodes During Charge/Discharge

#### 4.3.1 Studies on E-MAM

In the realm of MAM, comprehending the intricate mechanics that unfold during charge and discharge cycles holds pivotal significance. A compelling exploration lies within the phenomenon of charge/discharge-induced interfacial sliding, which is instigated beyond a critical stress threshold. The occurrence of fracture at the interface hinges upon the concentration profiles of intercalated atoms. Remarkably, studies in this domain remain relatively scarce, contributing to a dearth of insights into the mechanics governing MAM electrodes during these processes.

In the context of E-MAM, Basu et al.[56] undertook Grand Canonical Monte Carlo (GCMC) calculations to investigate interfacial stress dynamics during lithiation and delithiation. Focusing on two scenarios: (i) a-Si interfaced with a-Si, and (ii) a-Si interfaced with graphene, their findings unveiled lower interfacial stress during the cycling of the a-Si/graphene interface compared to the a-Si/a-Si counterpart. Remarkably, the performance of the a-Si/graphene interface surpassed that of non-slipping a-Si/Cu and a-Si/Ni interfaces (Fig. 6a) during cycling.

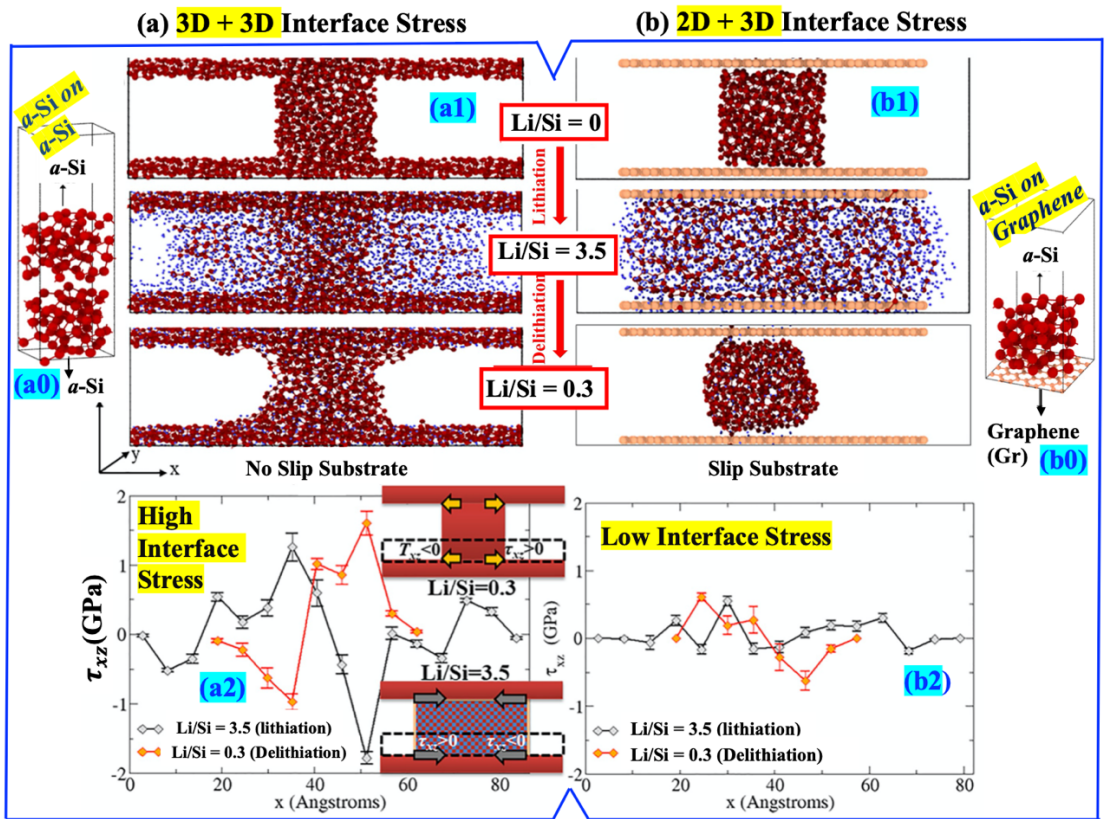

**Figure 6:** **(a0)** 3D+3D and **(b0)** 2D+3D **interfacial stress**: **(a1, b1)** Structural variation during charge/discharge, **(a2, b2)** Interface stress is high between Si, Si (3D+3D) but low between Si and graphene (3D + 2D). Reproduced with permission from Ref[56].

### 4.3.2 Studies on N-MAM

Turning to N-MAM, the mechanics of charge/discharge processes take on diverse complexities. In the case of NTO N-MAM, stress generation and fracture behavior are orchestrated by the crystallographic block structures. These structures facilitate lattice rearrangement, mitigating the impact of volume expansion. In the context of MVO N-MAM, the nanoporous geometry and tunnel shape play pivotal roles in influencing mechanical stability, kinetics, and thermodynamics. Hence, a thorough analysis of the interplay between crystal blocks, nanopore geometry, and tunnel shape is essential in understanding the electro-chemo-mechanical behaviors of N-MAM.

Kocer et al.[63, 64] delved into the world of cation disorder and lithium insertion mechanisms within Wadsley-Roth crystallographic shear phases through first-principles investigations. Fig. 7 exemplifies the lithiation of the $Nb_{12}WO_{33}$ structure, accompanied by lattice contraction along specific crystal directions that effectively buffer volume expansion. Their work underscores the evolution of localized and long-range electronic structures during lithiation, contributing to the enhanced performance of these materials as battery electrode. These insights, originating from the intrinsic crystallographic shear structure, likely hold relevance for various crystallographic shear phases involving niobium-titanium oxide or pure niobium oxide.

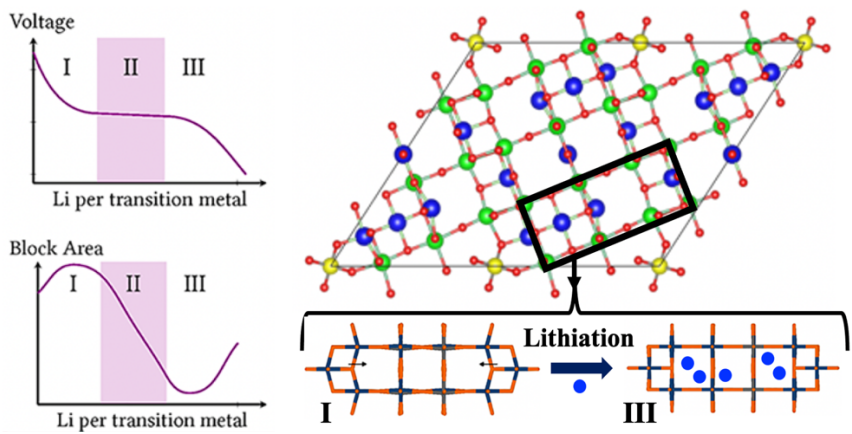

**Figure 7:** $Li_xNb_{12}WO_{33}$: Structural change of Wadsley-Roth crystallographic block upon lithiation (DFT study). Reproduced with permission from Ref[63].

In the intricate realm of MVO N-MAM, the interplay between tunnel shapes and intercalation mechanisms assumes paramount importance. As exemplified in Fig. 8, the insertion of Ca into the rectangular tunnel of tetragonal MVO ($MoVO_5$) leads to extensive structural distortion. In contrast, the insertion of Ca into a pentagonal tunnel induces no such distortion. Hence, a meticulous study of tunnel shapes is imperative when evaluating the intercalation mechanism and electrode stability of MVO for specific multivalent-ion-based energy storage applications.

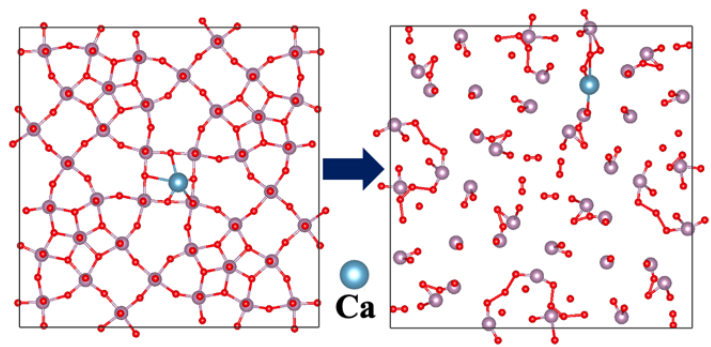

**Figure 8:** Chemo-mechanical instability of tetragonal MVO upon Ca intercalation at rectangular tunnel (DFT study). Work in progress by the author.

#### 4.4 Progress on Machine Learning Modeling of MAM Electrodes

The landscape of E-MAM is characterized by a plethora of possibilities arising from the diversity in nanoparticle (NP) shapes, sizes, and materials, as well as the intricate combinations of 2D materials with various NPs. Similarly, the realm of N-MAM boasts a myriad of potential compositions[65] and stoichiometries, presenting a challenge akin to finding a "needle in a haystack". Unveiling the optimal MAM configuration within this complexity is akin to solving a complex puzzle. As elucidated, atomistic studies prove computationally expensive, thus rendering the development of a machine learning framework for rapid prediction of MAM properties an urgent necessity. However, the scarcity of training data currently hampers the progress of machine learning initiatives in this direction.

Recently, Sharma et al.[66] developed MHDNN - *Modified High-Dimensional Neural Networks* (Figs. 9,10). Fig. 9 shows MHDNN model, trained by computationally expensive DFT data to develop machine-learned *Potential Energy Surfaces* (PES). However, this model considers only "*short-range energy*", *i.e.*, interaction between atoms closer to each other than a cutoff radius $R_c$, irrespective of physical nature of interactions. *Atom-centered symmetry function* (ACSF) represents structural information. The atomic interactions are *described by the local chemical environments instead of using a single neural network for the global PES.* Accordingly, the sum of atomic energy ($E_j^i$) contributes to total potential enegy,

$$E_{short} = \sum_{i=1}^{n} \sum_{j=1}^{N_i} E_j^i \tag{1}$$

Here, $n$ is the number of elements in the system, and $N_i$ is the number of atoms of element $i$ (Fig. 9). In MHDNN (Fig. 9b), the weights and architecture of all *Atomic Neural Networks* (ANN) are same. This ensures the invariance of total energy against interchanging of atoms within the network. Thus, MHDNN permits easy size extrapolation if new atoms are added.

**Example Problem:** Sharma et al.[66] implemented MHDNN to study formation energy of Sn-graphene interface (Fig. 10). The MHDNN was trained by equilibrium energy of several structures: *single and bilayer graphene*, *bulk Sn allotropes*, *few Sn atoms on graphene surface*, *bulk Sn with different phases (α, β)*. The machine learning predicted ($E_{predict}$) and DFT computed ($E_{DFT}$) energy agrees reasonably well (Fig. 10).

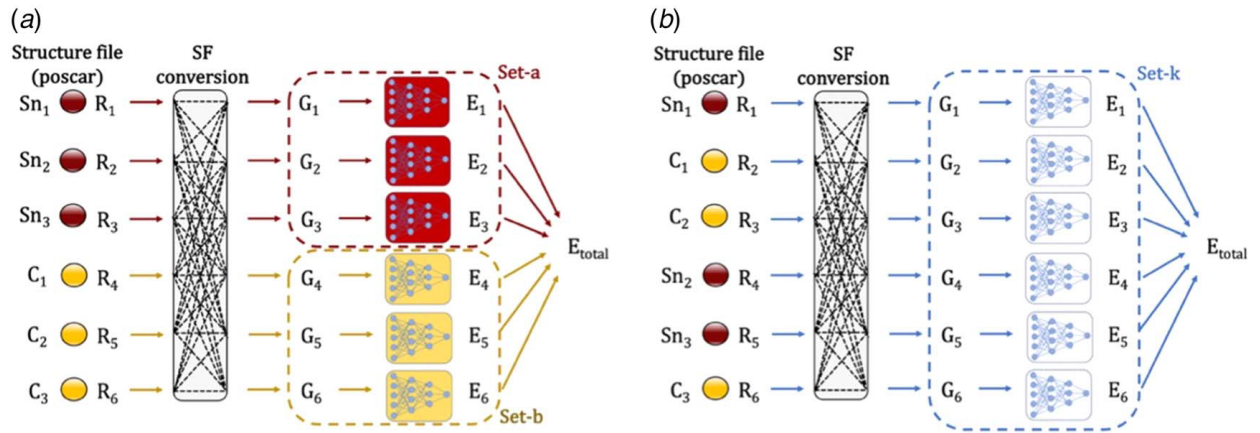

**Figure 9:** Comparative schematics of HDNN for bicomponent (Sn|C) system: **[a]** HDNN by Behler and Parrinello (BPNN) for bicomponent systems where weights and architecture of atomic neural networks (ann) are the same for single chemical species. Red-ann in set-a corresponds to Sn atoms and yellow-ann in set-b corresponds to C atoms and **[b]** Modified HDNN in the present study for bicomponent systems. Weights and architecture of all atomic neural networks (ann) are the same and correspond to the Sn|C system rather than single species. Atomic species are differentiated by the added feature of atomic number. Reproduced with permission from Ref[66].

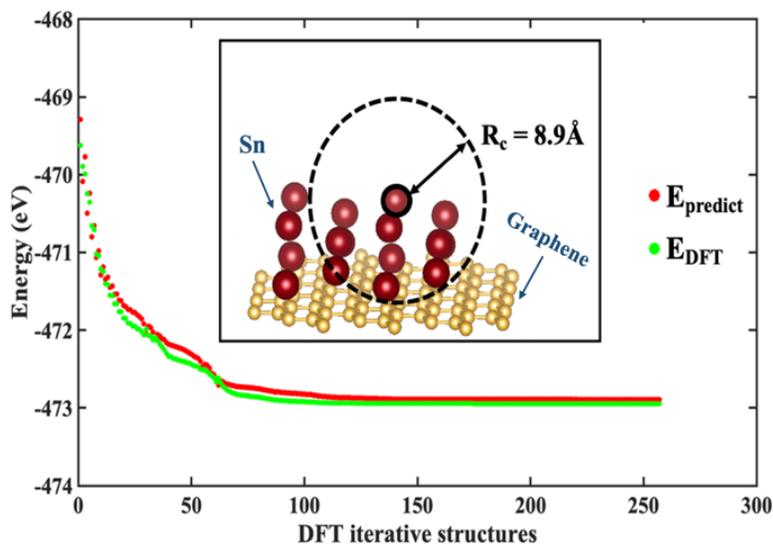

**Figure 10:** Test structures Sn over graphene. $E_{predict}$ and $E_{DFT}$ are total energies predicted by MHDNN and DFT, respectively. The dashed sphere with cut off radius $R_c = 8.9$Å represents chemical neighborhood that was observed for all atoms in the system. Reproduced with permission from Ref[66].

While strides have been made in MAM through machine learning, challenges persist due to the intricacies of computational endeavors. In the broader context of Materials Modeling, the work by Sharma et al.[66] remains a pioneering effort, particularly in the domain of interfaces. This underscores the vast untapped potential within this realm. In the subsequent section, we delve into some of the open problems that demand exploration and innovation.

## 5. MODELING OF MAM: OPPORTUNITIES AND CHALLENGES

The landscape of MAM is rife with intricate challenges and uncharted territories. The potential for groundbreaking discoveries is vast, yet the complexities involved necessitate focused efforts and innovative approaches. Here, we outline a few key challenge and open questions that beckon exploration in the realm of MAM.

Table 1 summarizes some potential MAM systems that can be studied to get started in this completely open field.

| E-MAM | | N-MAM |
|---|---|---|
| **Nanoparticles** | **2D Materials** | *Collect crystal structures from existing database* |
| **Si** - Silicon<br>**Se** - Selenium<br>**Sn** - Tin<br>**S** - Sulfur<br>**Ge** - Germanium | **Graphene :** pristine and defective<br>**TMD :** $MoS_2$, $WS_2$<br>**MXene:** $Ti_2C$, $Ti_3C_2$<br>*(different termination)* | **Niobium Tungsten Oxides (NTO) :** $Nb_{12}WO_{33}$, $Nb_{14}W_3O_{44}$, $Nb_{16}W_5O_{55}$, ......., *etc.*<br>**Molybdenum Vanadium Oxides (MVO):** Orthorhombic $(MoV_2O_8)$, Trigonal $(MoV_3O_6)$, Tetragonal $(MoVO_5)$, ........, *etc.* |
| **NP + NP :** Si+Si, Se + Se, .....<br>**NP + 2D materials :** Si + graphene, Sn + $MoS_2$, Se + $Ti_2C$, .......... | | |
| Different NP size, shape (orientation) | | |
| **Ion Intercalation Type : Only** $Li^+$ | | **Ion Type :** $Li^+$, $Ca^{+2}$ |
| **Electrolytes ::: Non-aqueous :** EC, DMC ; **Aqueous :** $LiTFSi \cdot H_2O$ | | |

**Table 1:** Different systems and parameters that can be studied

## 5.1 Interface Mechanics of *E*-MAM Electrodes

### 5.1.1 Interface Adhesion Study

The adhesion between interfaces is a critical factor in determining the stability and performance of MAM systems. To assess the interface adhesion, a fundamental step is computing the work of separation ($W_{sep}$), as outlined in Equation 2, for various NP+NP and NP+ 2D materials systems (Table 1, Fig. 11). This computation aids in quantifying the energy required to separate two materials from the interface, revealing insights into their adhesive properties. A closer look at the equation[56] and considerations for its calculation can provide valuable insights:

$$W_{sep} = S_1 + S_2 - \gamma_{12} = (E_1^{tot} + E_2^{tot} - E_{12}^{tot})/A \qquad (2)$$

Where $S_i$ is the surface energy of slab $i$ ($i = 1, 2$), $\gamma_{12}$ is the interface energy, $E_i^{tot}$ is the total energy of material $i$ ($i = 1, 2$). $E_{12}^{tot}$ is the total energy of the combined system. $A$ is the interface area.

In practical terms, this equation encapsulates the energy associated with detaching two materials from each other across an interface. The greater the required energy, the stronger the adhesive interaction between the materials. Notably, implementing DFT[67, 68] calculations using vdW-inclusive Generalized Gradient Approximation (GGA)[69, 70] functionals is a powerful approach for computing $W_{sep}$ in various NP+NP and NP+2D systems. As demonstrated in Fig. 4a, this process involves determining the total energy of individual materials ($E_1^{tot}$, $E_2^{tot}$) and the combined system ($E_{12}^{tot}$), factoring in the surface energies ($S_1$, $S_2$) and interface energy ($\gamma_{12}$).

Sharma et al.[58] explored this concept in their work, revealing that factor like intercalation (e.g., lithiation) do not significantly influence $W_{sep}$. This implies that variations in interface adhesion due to intercalation may not be a critical concern. Instead, focus can be directed toward understanding and controlling the intrinsic adhesion between different materials at interfaces.

The analysis of interface adhesion serves as pivotal cornerstone in designing robust MAM systems, guiding the selection of materials and configurations that can withstand the demands of electro-chemo-mechanical processes. By mastering the intricacies of interface adhesion, researchers can contribute significantly to the advancement of energy storage technologies and related fields.

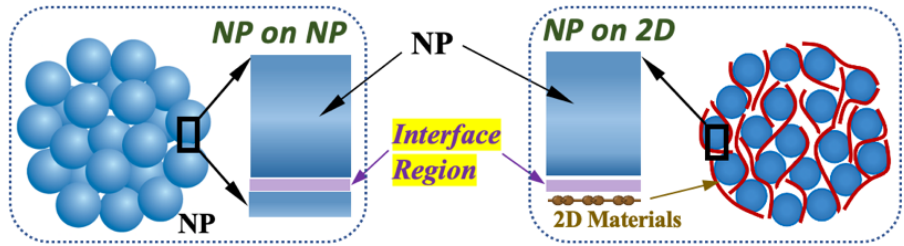

**Figure 11.** In *E*-MAM, NP-NP and NP-2D materials interface.

### 5.1.2 Correlating Potential Gradient (φ) with $W_{sep}$

Sharma et al.[58] shed light on a fascinating correlation between the potential gradient (φ) and $W_{sep}$ for interfaces involving Se/graphene and Si/graphene systems (Fig. 4b2). This innovative analysis holds the potential to unveil deeper insights into the interplay between electrostatic and adhesive properties at interfaces. The correlation hinges on the gradient of the electrostatic potential (φ), which is computed perpendicular to the interface. This gradient, represented as $d\varphi/dz$, is calculated using the formula:

$$d\varphi/dz = \left(U_{Se} - U_{graphene}\right)/d \qquad (3)$$

Here, $U_{Se}$ and $U_{graphene}$ denote the electrostatic potentials of Se and graphene, respectively, obtained through DFT calculations. $d$ signifies the interface gap.

The correlation between φ and $W_{sep}$ opens doors to a deeper understanding of how electrostatic forces influence adhesive properties. The connection between these two factors provides a novel perspective on interface behavior, potentially guiding material selection, and design considerations. This methodology could potentially be extended to diverse systems, investigating the correlation between φ and $W_{sep}$ for a range of interfaces.

By revealing the intricate interplay between electrostatic potential and interface adhesion, this correlation empowers researchers with a valuable tool for predicting and optimizing the performance of MAM systems. As scientists further explore this avenue, new insights could emerge, ultimately advancing the development of energy storage technologies.

### 5.1.3 Interfacial Charge Transfer and Bonding Character

The analysis of interfacial charge transfer ($\rho$) and bonding character in MAM interfaces offers a profound insight into the intricacies of their electrochemical behavior and structural stability. Recent work by Sharma et al.[58] (Fig. 4b1) highlighted the significance of Bader charge transfer in understanding interface dynamics. Extending these analyses to various interfaces could unravel further complexities and correlations, enriching our understanding of MAM systems.

The assessment of interfacial charge transfer[60, 61] ($\rho$) involves quantifying the exchange of electrons across the interface. This process not only influences the electronic structure but also affects the adhesive properties and electrochemical behavior of MAM interfaces. To comprehend the relationship between $\rho$ and adhesive properties, similar analyses, like those demonstrated by Sharma et al.[58] can be conducted across diverse interfaces. The subsequent correlation of $\rho$ with the $W_{sep}$ can shed light on the role of $\rho$ in interface adhesion. Unraveling the bonding character at MAM interfaces is equally pivotal. The Electron Localized Function (ELF)[71] offers a powerful tool for mapping the distribution of electron density[60], revealing insights into the nature of chemical bonds and their strength across the interface. Employing ELF analysis across various interfaces can provide a nuanced understanding of interfacial bonding. This information is vital for predicting the chemo-mechanical stability, reactivity, and durability of MAM systems.

These detailed analyses hold transformative potential. By combining insights from interfacial charge transfer and bonding character, researchers can identify E-MAM configurations with favorable $W_{sep}$, marking them for further investigation. Notably, the absence of such interfaces in N-MAM simplifies this aspect.

## 5.2 Mechanics of MAM Electrode-Electrolyte Interface and SEI Formation

### 5.2.1 Chemo-Mechanical Stability Between *E*- and *N*-MAM Electrodes and Electrolytes

#### 5.2.1.1 Interphase Formation, Structural Stability, Interface Stress

Undesirable chemical reactions at the electrode/electrolyte interface can initiate chemo-mechanical degradation, compromising the performance of MAM systems. For instance, the dissolution of transition metals like Mn and Ni in organic liquid electrolytes is a recurring issue with many Mn-containing electrodes, leading to structural distortions[72, 73]. The consequences of these reactions are particularly pronounced at the interface, where structural integrity is compromised. Prior to delving into processes of intercalation and deintercalation (charge and discharge), it is imperative to assess the electrode stability itself. Evaluating electrode stability ensures that the electrochemical reactions occur under controlled conditions. Fig. 12a shows a setup for electrode/electrolyte interface stability calculations.

Classical MD simulations can provide insights into interface stability for systems where suitable interatomic potentials are available. However, for complex systems lacking appropriate potentials, Ab initio Molecular Dynamics (AIMD)[74, 75] calculations offer an alternative. AIMD simulations delve into quantum-level interactions, but their feasibility decreases with system size due to computational challenges. Chemical reactions between electrodes and electrolytes generate interphase products at the interface. In cases of unstable electrodes, the width of these interphase products ($d$) continues to increase over time. To comprehensively understand the implications of these interphases, detailed analysis[76] of their composition width, and stress[77] ($\sigma_{int}$) is essential. Identifying the critical stress levels at which interphases undergo rapid expansion is pivotal, as this threshold signifies the point at which degradation becomes prominent.

Through this systematic analysis, researchers can gain insights into the interplay between interphase formation, structural stability, and mechanical stress at electrode-electrolyte interfaces. Armed with this knowledge, they can make informed decisions regarding material selection, electrode design, and the development of protective layers to enhance the overall performance and durability of MAM systems.

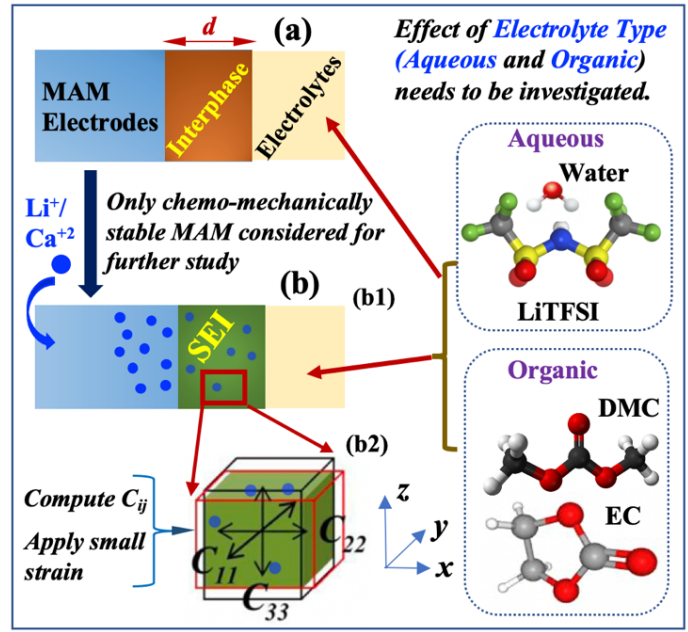

**Figure 12: [a]** Interphase formation between MAM electrode and electrolyte. **[b]** (b1) SEI formation, (b2) SEI mechanical properties computation.

### 5.2.1.2 Correlation of Interphase Formation Energy ($E_f$) and Maximum Interface Stress ($\sigma_{int}$)

Understanding the correlation between the interphase formation energy[40, 41, 50] ($E_f$) and maximum interface stress ($\sigma_{int}$) provides deeper insights into the driving forces behind interphase growth and structural stability. The $E_f$ is a pivotal parameter that characterizes the energy required to create interphase products at the electrode/electrolyte interface. The $E_f$ can be computed with DFT for all cases as:

$$E_f(d) = E(d) - E_{electrolyte} \qquad (4)$$

Here, $E(d)$ and $E_{electrolyte}$ are energies of interphase and electrolytes, respectively. This value signifies the thermodynamic stability of the interphase in the given system. Maximum electrode/electrolyte interface stress ($\sigma_{int}$) refers to the highest stress due to chemo-mechanical interactions. This stress can arise from factors such as volume changes, chemical reactions, and interphase growth. Identifying the critical threshold at which $\sigma_{int}$ triggers interphase growth and structural deterioration is a key aspect of ensuring electrode integrity.

By correlating $E_f$ with $\sigma_{int}$, researchers can glean insights into the interplay between thermodynamics and mechanics at the electrode/electrolyte interface. A strong correlation might suggest that higher interphase formation energies correspond to higher interface stresses. This insight could indicate that energetically favorable interphase formation is associated with increased structural strain. Conversely, a weak or negative correlation could indicate that the interphase formation is driven by mechanisms other than mechanical stress. This could encompass electrochemical factors, interface charge transfer, or dynamic interfacial processes. In summary, the correlation of $E_f$ and $\sigma_{int}$ adds a new layer of understanding to the behavior of MAM systems at the electrode/electrolyte interface. This correlation could help elucidate the complex interplay between thermodynamics stability and mechanical stress, ultimately informing strategies for designing stable and long-lasting energy storage devices.

### 5.2.2 Mechanics of SEI Formation and its Mechanical Properties

In the journey of understanding MAM electrode-electrolyte interfaces, the mechanics of SEI formation takes the center stage. This aspect is critical for stable MAM systems, as described in Section 5.2.1. Investigating SEI formation during charge and discharge stages offers valuable insights into the behavior of these materials.

#### 5.2.2.1 Mechanical Properties of SEI

SEI formation during intercalation and deintercalation stages is a dynamic process that significantly impacts the performance of energy storage systems. To explore this phenomenon, researchers often mimic different charge states by introducing various percentages of intercalated atoms near the interface region (Fig. 12b). Evaluating the mechanical properties of SEI involves calculating its elastic constants ($C_{ij}$), which provide insights into its mechanical compliance. Through MD or AIMD simulations, researchers can study the SEI region (as extracted in Fig. 12b) to compute these elastic constants. By subjecting the SEI to small strains along the coordinated axes, stress changes can be analyzed to derive the elastic constants.

From the elastic constants, bulk ($B$) and shear ($G$) moduli can be computed, providing essential information about the SEI's stiffness and mechanical behavior. The bulk modulus (B) quantifies the material's response to hydrostatic compression, while the shear modulus (G) characterizes its response to shear stress. These properties are pivotal in understanding how the SEI responds to mechanical deformation.

Bulk ($B$) and shear ($G$) modulus can be computed as[62] –

$$B = \left((C_{11} + C_{22} + C_{33}) + 2(C_{12} + C_{13} + C_{23})\right)/9 \tag{5}$$

$$G = \left((C_{11} + C_{22} + C_{33}) - (C_{12} + C_{13} + C_{23})\right)/15 + (C_{44} + C_{55} + C_{66})/5 \tag{6}$$

Young's modulus can be computed by assuming isotropic linear elastic stress-strain relations. These calculations will reveal whether SEI will be compliant or stiffer than active materials. Moreover, these computed properties can be used in future continuum modeling.

### 5.2.2.2 Correlation of SEI Formation Energy ($E_f^{SEI}$) and its Mechanical Properties

SEI formation energy can be computed with DFT as:

$$E_f^{SEI} = \left(E_{SEI} - \left(E_X + E_{electrolyte}\right)\right)\Big/ N_X^{SEI} \tag{7}$$

Here $X$ = intercalated atoms. $N_X^{SEI}$ is the number of $X$ atoms that have reacted to form the SEI. For different $X$ atoms, $E_f^{SEI}$ can be computed and correlated with the SEI mechanical properties. Correlating $E_f^{SEI}$ with the mechanical properties of SEI offers a deeper understanding of how interfacial chemistry impacts mechanical behavior. This correlation can provide insights into the relationship between SEI formation energy and its response to mechanical stress. Such insights are invaluable for optimizing MAM systems, improving their stability, and guiding material design strategies.

## 5.3 Mechanics of MAM Electrodes During Charge/Discharge

### 5.3.1 Study of Fracture, Interface Stress, and Sliding During Charge/Discharge in *E*-MAM

As we delve into the mechanics of E-MAM during the charge and discharge cycles, the focus shifts towards understanding fracture mechanisms, interface stress, and sliding behavior.

### 5.3.1.1 Interfacial Stress During Charge/Discharge

Simulating the interfacial stress that arises during charge and discharge processes provides crucial insights into the stability and mechanical behavior of E-MAM systems. Grand Canonical Monte Carlo (GCMC)[56] simulations, facilitated by software like LAMMPS[78], can be employed to compute interfacial stress under varying charge/discharge rates. Building on the work of Basu et al.[56], who used ReaxFF[77, 79, 80] potentials for stress analysis[56], suitable interatomic potentials can be utilized for various interfaced systems.

The ***charging*** can be modeled in three-step process[56] – ***(i)*** *intercalated atoms (e.g., Li) can be introduced in regions defined close to NP surfaces,* ***(ii)*** *followed by application of a small external force to these intercalated atoms to drive them into NP in a process analogous to the voltage-driven intercalation in an electrolyte cell[81],* ***(iii)*** *the freshly intercalated system is allowed to relax.* These steps can be repeated for electrode intercalation to the desired Li content. ***Discharging*** process can be carried out by selecting intercalated atoms closest to NP surfaces, which will be then pulled out by applying an external force in a direction opposite to charging.

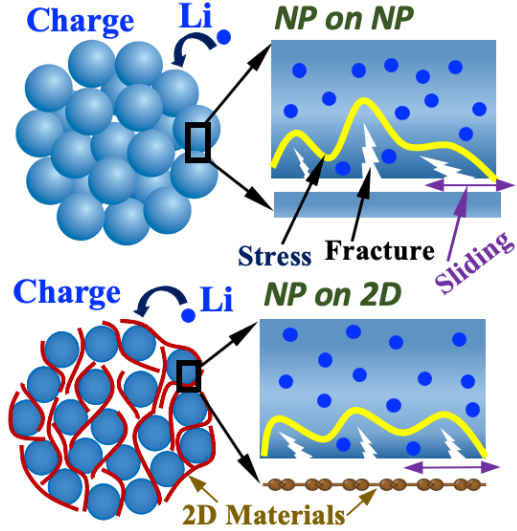

**Figure 13:** Interfacial stress, sliding, fracture in *E*-MAM

### *5.3.1.2 Interface Sliding During Charge/Discharge*

Investigating interface sliding is pivotal for understanding the dynamics of electrode-electrolyte interactions. The interfacial traction during sliding is the derivative of $W_{sep}$ with respect to the sliding distance $l$ and can be computed from $W_{sep}$ as a function of intercalation content (*e.g.*, Li content $x$ in Li$_x$Se, $x = 0$ to 1) according to[82] –

$$\frac{\partial \gamma_{12}}{\partial l} = -\frac{\partial W_{sep}}{\partial l}\Big|_x \tag{8}$$

The critical interface shear stress to initiate sliding is thus the maximum of $\frac{\partial \gamma_{12}}{\partial l}$ or $\tau_{max} = \frac{\partial \gamma_{12}}{\partial l}\Big|_{max}$. DFT calculations can be performed to obtain $W_{sep}$ profile and determine the $\frac{\partial \gamma_{12}}{\partial l}$ profile along the sliding. This analysis allows researchers to gauge the conditions under which sliding behavior is triggered.

### *5.3.1.3 Fracture Mechanisms During Charge/Discharge*

Understanding crack initiation and propagation (Fig. 13) during charge and discharge stages is essential for ensuring the structural integrity of MAM electrodes. The interplay between interfacial stress, sliding behavior, and fracture can provide insights into failure mechanisms. Through detailed analyses and correlation with interface stress, researchers can pinpoint the conditions under which fractures are likely to occur. By exploring these aspects, researchers aim to unravel the complexities of mechanical behavior during charge and discharge cycles in E-MAM systems. This understanding not only enhances the durability of energy storage devices but also informs strategies for materials design and electrode engineering to mitigate fracture risks and optimize performance.

### 5.3.2 Study of Stress, Fracture, and Volume Change During Charging of *N*-MAM

This section delves into the intricate dynamics of stress, fracture, and volume change within N-MAM during charging. N-MAMs, particularly those structured with complex oxides like Nb-W-O and Mo-V-O, offer unique insights and challenges. For most *N*-MAMs, suitable interatomic potentials are unavailable yet for GCMC study. Therefore, different charging states can be mimicked by AIMD calculations considering various percentages of $X$ ($X$ = monovalent (e.g., $Li^+$) or multivalent ions (e.g., $Ca^{+2}$) having one- and two-oxidation states, respectively).

### 5.3.2.1 Study of NTO Structures

Block-type crystal structures with Wadsley-Roth crystallographic shear phases are key elements forming the NTO structures[11] (Fig. 2).

#### 5.3.2.1.1 Intercalation Mechanism, Structural Distortion, Voltage Correlation

Kocer et al.'s work[63], as shown in Figure 7, provides valuable insights into structural changes during lithiation in $Nb_{12}WO_{33}$. AIMD calculations can be harnessed to mimic different charging states by varying the percentages of $X$ (monovalent or multivalent ions) at different locations within the structure.

The maximum $X$ content corresponds to negative voltage $V$. The $V$ will be computed[40, 50] as $V = -\Delta E_f/n$. Here, $E_f$ is the formation energy to be obtained by DFT. $n$ is the number of intercalated $X$. For varying $V$ (*i.e.*, $X$ contents), it is important to study how local structural distortions lead to the lattice contraction along specific crystallographic directions, *buffering the volume expansion of the materials.* For a deeper analysis of structural distortions, following the procedure by Kocer et al.[63], for different intercalated NTO structures, three *distortion measures* need to be studied: *(i) a dimensionless bond angle variance* $\Delta(\theta_{oct})$, *(ii) the quadratic elongation* $\lambda_{oct}$, and *(iii) an off-centering distance* $d_{oct}$. For different $X$, plotting $V$, $\Delta(\theta_{oct})$, $\lambda_{oct}$, $d_{oct}$ will provide insight into the structural distortion at different charging states.

#### 5.3.2.1.2 Stress and Fracture

Employing DFT calculations, researchers can analyze stress by applying strain[83] at various charging states. This allows for the investigation of stress distribution and its effects on fracture initiation and propagation. By identifying locations of crack initiation and understanding how fracture propagate, researchers gain insights into the material's mechanical response during charging.

### 5.3.2.2 Study of MVO Structures

The complexities of MVO (Molybdenum Vanadium Oxides) structures[14] add another layer to the investigation. Different polymorphs (orthogonal, trigonal, tetragonal) with varying tunnel shapes (rectangular, pentagonal, hexagonal, heptagonal) pose intriguing challenges. Similar to NTO structures, analyses of structural distortion, voltage correlation, stress distribution, and fracture mechanics need to be performed for MVO structures. Fig. 8 shows complete structural degradation upon Ca intercalation inside rectangular tunnel in tetragonal MVO. All activities mentioned for NTO structures need to be performed for various MVO (Fig. 14a) polymorphs.

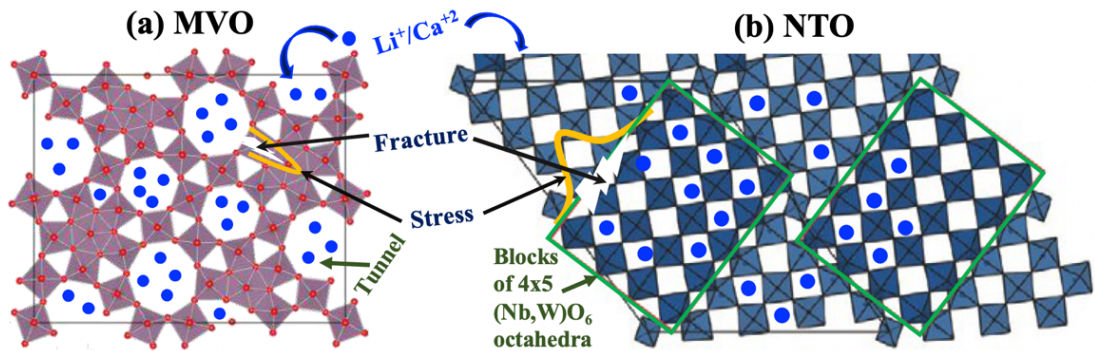

**Figure 14:** Chemo-mechanics of **[a]** MVO and **[b]** NTO during charging

## 5.4 Data-Driven Mechanics of MAM Electrodes

The emerging field of data-driven mechanics for MAM electrodes holds promise for accelerating material discovery and optimization. Building on the foundational work by Sharma et al.[66], this section outlines the need to extend and enhance data-driven approaches in understanding the interfacial properties and behaviors of MAM electrodes during charging and discharging processes.

A long-term question is – for E-MAM, can we develop a machine learning framework that will take various 2D/3D combination structures, intercalation atoms, and predict at different intercalation stages, the interface adhesion, interfacial stress, and interfacial charge transfer (Fig. 15)? Similarly, can we develop machine learning framework for predicting various properties of N-MAM ?

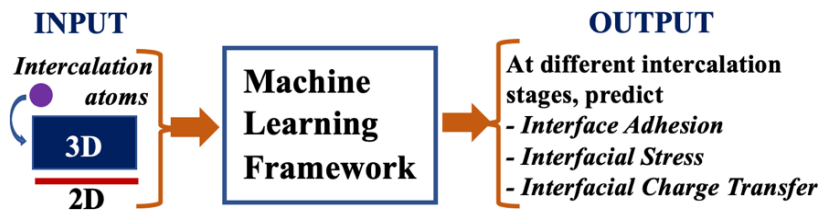

**Figure 15:** Outlook: For E-MAM, a machine learning framework that can take various combinations of 2D/3D systems, intercalation atoms and predict different interfacial properties such as interface adhesion, stress, charge transfer, etc. Similarly, we need machine learning framework for N-MAM.

### 5.4.1 Need for More Training Data for Interface Formation Prediction

The pioneering efforts by Sharma et al.[58] in predicting interface formation energy for Sn/graphene interfaces demonstrated the potential of machine learning in predicting material properties. However, the scope needs to be broadened. Besides E-MAM, machine learning framework for N-MAM (e.g., naturally occurring oxide materials) is also urgently necessary. A comprehensive training dataset needs to be generated, primarily thorough DFT calculations. This expansive dataset will enable machine learning models to accurately predict various required properties of MAM across a diverse range of material combinations.

### 5.4.2 Implementation of Force and Stress Computation in MHDNN

The incorporation of force and stress computations into machine learning (ML) models represents a crucial step toward deeper insights. There can be different approaches for implementation of force and stress in machine learning modeling. The approach described by Behler et al.[84, 85] serves as a guide for this implementation. The force and stress implementation in ML models is essential for understanding how stress affects the structural integrity and performance of MAM electrodes during charge and discharge cycles.

### 5.4.2.1 Force calculation

The force component $F_{\beta p}$ acting on atom $\beta$ in direction $p = \{x, y, z\}$ can be computed[86] by the negative derivative of $E_{\text{short}}$ (Eq. 1) with respect to coordinate $R_{\beta p}$ as:

$$F_{\beta p}^{\text{short}} = -\frac{\partial E_{\text{short}}}{\partial R_{\beta p}} = -\sum_{i=1}^{N_{\text{elem}}} \sum_{j=1}^{N_{\text{atom}}^i} \frac{\partial E_j^i}{\partial R_{\beta p}} = -\sum_{i=1}^{N_{\text{elem}}} \sum_{j=1}^{N_{\text{atom}}^i} \sum_{k=1}^{N_{sym}^i} \frac{\partial E_j^i}{\partial G_{jk}^i} \frac{\partial G_{jk}^i}{\partial R_{\beta p}} \tag{9}$$

$N_{sym}^i$ is the number of symmetry functions used to describe the local environments of atoms of element $i$.

### 5.4.2.2 Stress calculation

The stress tensor $\boldsymbol{\sigma}$ contains a kinetic and static contribution[86], $\sigma = \sigma^{\text{kinetic}} + \sigma^{\text{static}}$. The kinetic stress contribution ($\sigma^{\text{kin}}$) is a dynamical property that will be computed from the atomic velocities $v_i$ and the atomic masses $m_i$. Its components can be computed as –

$$\sigma_{mn}^{\text{kin}} = \frac{1}{V} \sum_{k=1}^{N_{atom}} m_k v_{kp} v_{kq} \tag{10}$$

with $p$ and $q$ being $\{x, y, z\}$, and $v_{kp}$ and $v_{kq}$ being components of the velocity vector. $V$ is the volume of the simulation cell. The static stress $\sigma^{\text{static}}$ depends on the atomic positions and can be calculated from the forces. Defining the cartesian coordinate difference as $R_{ij,p} = R_{i,p} - R_{j,p}$, the contributions of the radial and angular symmetry functions for $i$ will be determined separately according to

$$\sigma_{i,pq}^{\text{static,rad}} = \sum_{j=1}^{N_{atom}} R_{ij,p} F_{jq} = -\sum_{k=1}^{N_{atom}} \sum_{\beta=1}^{N_{sym}^k} \frac{\partial E_k}{\partial G_{k\beta}} \sum_{j=1}^{N_{atom}} R_{ij,p} \frac{\partial G_{k\beta}}{\partial R_{jq}} \tag{11}$$

with $F_{jq}$ being the force component of atom $j$ in direction $q$. The angular stress contribution of atom $i$ is

$$\sigma_{i,pq}^{\text{static,ang}} = \sum_{j=1}^{N_{atom}} R_{ij,p} F_{jq} + \sum_{k=1}^{N_{atom}} R_{ik,p} F_{kq} = -\sum_{k=1}^{N_{atom}} \sum_{\beta=1}^{N_{sym}^k} \frac{\partial E_k}{\partial G_{k\beta}} \left( \sum_{j=1}^{N_{atom}} R_{ij,p} \frac{\partial G_{k\beta}}{\partial R_{jq}} + \sum_{r=1}^{N_{atom}} R_{ir,p} \frac{\partial G_{k\beta}}{\partial R_{rq}} \right) \tag{12}$$

For the evaluation of elements of stress tensor, the intermediate transformation of the cartesian coordinates to ACSFs will be considered in the force components. The complete static stress tensor elements can be obtained by summing over all atomic contributions

$$\sigma_{pq}^{\text{static}} = \sum_{i=1}^{N_{atom}} \left( \sigma_{i,pq}^{\text{static,rad}} + \sigma_{i,pq}^{\text{static,ang}} \right) \tag{13}$$

### 5.4.3 Implementation of Non-Local Long-Range Charge Transfer (NLLRCT) in MHDNN

Most ML models in electro-chemo-mechanics focus on short-range charge transfer. An atomistic study[87] (Fig. 16) demonstrated that the atomic charge at one end is different due to different surface termination on another end. Consider the effect of 'neighboring atoms' only (short-range) results in inaccurate prediction of charge (Q). Thus energy (E), force (F), and stress ($\sigma$) prediction is inaccurate since E, F, $\sigma$ = f(Q). Therefore, to achieve more precise results, it becomes necessary to incorporate long-range charge transfer (Fig. 17).

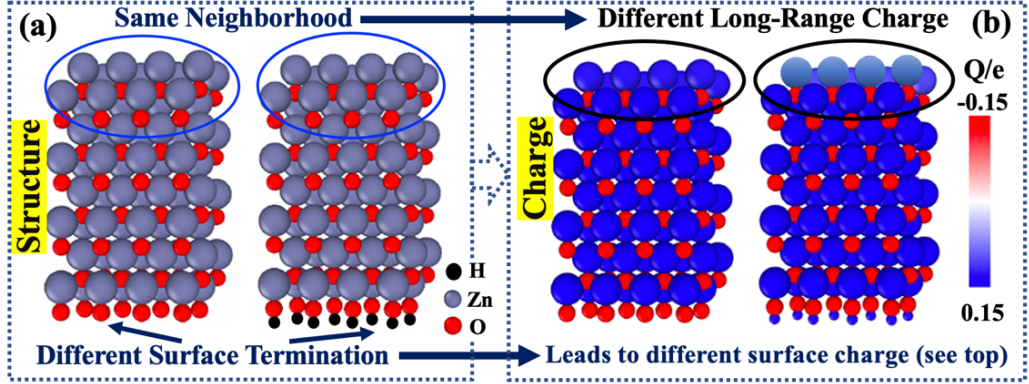

**Figure 16. Example of Long-range charge-transfer. [a]** Different surface terminations (*see bottom*) lead to **[b]** different surface charge (*see top*). Reproduced with permission from Ref[87].

Behler et al.[88] described the procedure of implementing NLLRCT. Fig. 17 shows the overall structure of MHDNN with NLLRCT for a system with *n* elements. The total energy consists of a *short-range part* and an *electrostatic long-range part*[85] –

$$E_{total}(\mathbf{R}, \mathbf{Q}) = E_{elec}(\mathbf{R}, \mathbf{Q}) + E_{short}(\mathbf{R}, \mathbf{Q}) \tag{14}$$

The electrostatic part $E_{elec}(\mathbf{R}, \mathbf{Q})$ depends on a set of atomic charges $\mathbf{Q} = \{Q_i\}$, which will be trained to reference charges from DFT calculations, and the atomic positions $\mathbf{R} = \{\boldsymbol{R}_i\}$. These charges will not be directly expressed by ANN as a function of the local atomic environment. Instead, they will be obtained indirectly from a **charge equilibrium scheme[89]** based on **atomic electronegativities** $\{\chi_i\}$ that will be adjusted to yield charges in agreement with the DFT reference charges. The $\chi$ are local properties defined as a function of the atomic environments. To predict the atomic charges, represented by Gaussian charge densities of width $\alpha_i$, taken from the covalent radii of the respective elements, **a charge equilibrium scheme[89] will be used**. The $E_{elec}$ will be determined based on the partial charges resulting from the fitted $\chi$. The short-range ANN will be trained to express the remaining part of the total energy $E_{ref}$ according to

$$E_{short} = E_{ref} - E_{elec} = \sum_{i=1}^{N_{atom}} E_i(\{\boldsymbol{G}_i\}, Q_i) \tag{15}$$

The electrostatic **force contribution** in direction $p = \{x, y, z\}$ will be computed as[85]:

$$F_p^{elec} = \sum_{j=1}^{N_{atom}} \sum_{i=1, i \neq j}^{N_{atom}} \frac{Q_i}{R_{ij}} \cdot \left[ \frac{1}{2} \frac{Q_i}{R_{ij}} \frac{\partial R_{ij}}{\partial p} - \sum_{k=1}^{N_{sym,j}} \frac{\partial Q_j}{\partial G_{jk}} \frac{\partial G_{jk}}{\partial p} \right] \tag{16}$$

The $F^{elec}$ (Eq. 16) will be added to $F^{short}$ (Eq. 9) to get the **net force**, *i.e.*, $F^{net} = F^{elec} + F^{short}$. With this $F^{net}$, $\sigma^{static}$ can be computed using Eqs. 11-13. The $\sigma^{net}$ can be obtained by adding $\sigma^{kin}$ (Eq. 9) to $\sigma^{static}$.

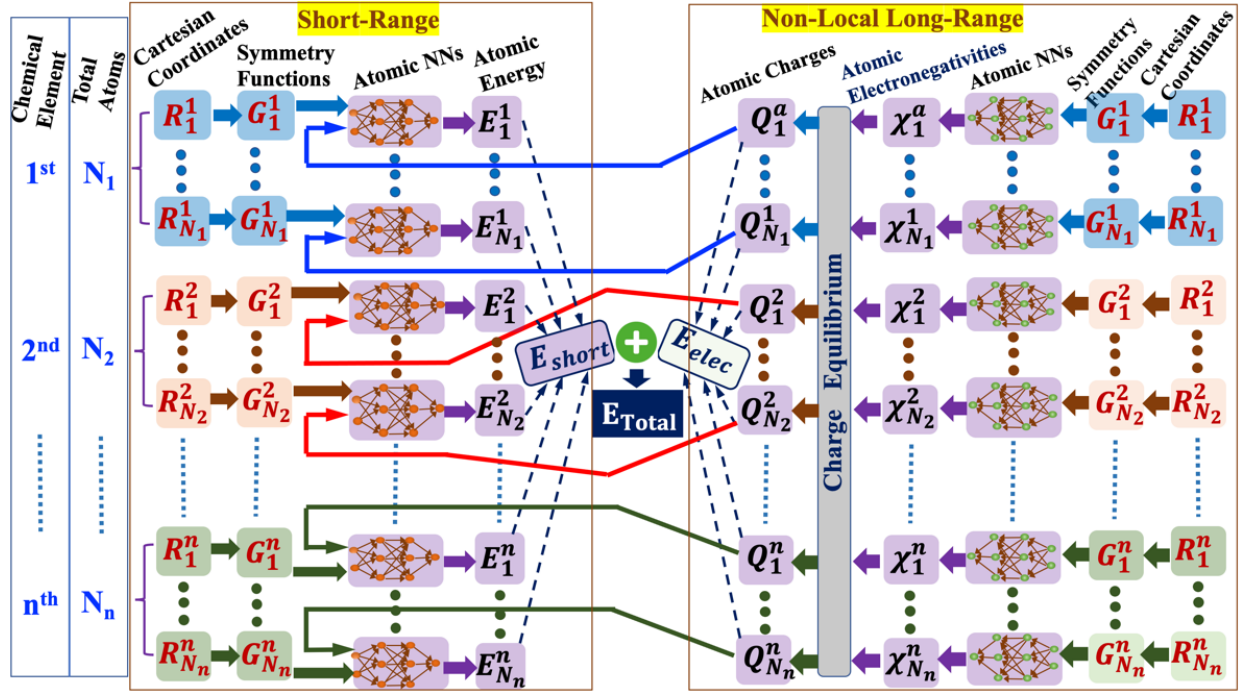

**Figure 17. Non-Local Long-Range Charge Transfer (NLLRCT) Implementation**

## 6. CHALLENGES IN MODELING MAM

The pursuit of modeling MAM brings forth several challenges that researchers must address to advance the understanding of these complex systems and their behavior in energy storage applications. The following challenges span different aspects of modeling, from computational methods to data availability.

### 6.1 Challenges in Density Functional Theory (DFT) Calculations

MAM systems often involve large and intricate structures, making DFT calculations computationally demanding and time-consuming. DFT calculations of interfaced structures are computationally very expensive. The complexity of crystal structures, the presence of multiple intercalation sites, and cation disorder in certain oxide materials pose significant challenges. These computational limitations hinder the comprehensive exploration of MAM systems through first-principles calculations. The need for high computational resources and the extended timeframes required for simulations can limit the scope of theoretical studies.

### 6.2 Challenges in Molecular Simulations

While MD simulations can handle larger systems, the availability of accurate and suitable interatomic potentials is a critical constraint. The ReaxFF potential was used by Basu et al. for Si/Si and Si/graphene systems but finding suitable potentials for other interfaced systems remains a challenge. This limitation particularly affects systems like N-MAM, where the lack of appropriate potentials has hindered exploration. The development of accurate interatomic potentials for diverse systems is essential for expanding MD simulations in MAM research.

## 6.3 Challenges with Continuum Modeling

Continuum modeling, while valuable for studying macroscopic behavior, falls short in providing atomic-scale insights into interfacial phenomena, charge transfer, and stress distribution. The challenge lies in bridging the gap between atomic-level understanding and macroscopic behavior to comprehensively capture the chemo-mechanical response of MAM materials.

## 6.4 Challenges in Machine Learning Modeling due to Lack of Training Data

Machine learning holds great potential in predicting material properties, but its effectiveness depends on the availability of reliable training data. Existing materials databases such as Materials Project[90, 91], OQMD[92], AFLOW[93], etc. often lack data for interfaced or intercalated systems, which are essential for MAM research. The work by Sharma et al.[66] serves as an example, but its applicability may be limited due to the scarcity of relevant training data. Generating a comprehensive database for MAM materials requires collaborative efforts within the scientific community to address the data gap.

Addressing these challenges will require interdisciplinary collaborative, innovative computational techniques, and the concerted efforts of researchers across fields such as materials science, computational chemistry, and machine learning. By overcoming these challenges, researchers can unlock the full potential of MAM materials for next-generation energy storage technologies.

## 7. CONCLUSIONS

The current battery industry predominantly utilizes microparticles as active materials, despite their inherent limitations. While nanomaterials offer various advantages, the industry hesitates to adopt them due to certain drawbacks. The concept of multiscale active particles, combining the benefits of both nano- and micro-worlds, presents a promising way forward. However, computational modeling in this direction remains relatively unexplored. This perspective has shed light on the need for MAM and outlined four major areas for computational investigation:

- **Interface Mechanics of Engineered MAM:** Understanding the interfacial behavior of E-MAM is crucial for their optimal design. Interfacial phenomena in composite electrodes, such as NP-NP and NP-2D assemblies, need to be studied to enable proper material selection and design. Interface adhesion, charge transfer, and potential gradient correlation are important factors to consider.

- **Mechanics of MAM Electrode-Electrolyte Interaction and SEI Formation:** The interaction between MAM electrode and electrolytes plays a vital role in battery performance. Investigating chemo-mechanical stability, SEI formation mechanisms, and mechanical properties of SEI products are essential for enhancing electrode stability and performance.

- **Mechanics of MAM Electrodes During Charge/Discharge:** The chemo-mechanical changes that occur in MAM electrodes during charge/discharge cycles introduces stresses, volume changes, and potential fracture. Studying these aspects is critical for preventing electrode failure and optimizing battery durability.

- **Data-Driven Mechanics of MAM Electrodes:** Developing machine learning frameworks for predicting properties of MAM electrodes is essential, given the computational complexity of these systems. However, the lack of training data is a major challenge that needs to be addressed for accurate predictions.

While progress has been made in these areas, significant computational challenges persist:

- **DFT Calculations:** DFT calculations for large MAM systems are computationally expensive, hindering their thorough exploration.

- **Molecular Simulations:** Stable interatomic potentials for MD simulations are lacking for many MAM systems, limiting their applicability.

- **Continuum Modeling:** Bridging the gap between atomic-level details and macroscopic behavior remains a challenge in continuum modeling.

- **Machine Learning with Limited Data:** Developing effective machine learning models for MAM materials requires a comprehensive database, which is currently lacking.

Addressing these challenges will require collaborative efforts from researchers across various disciplines. By tackling these open problems, the computational community can advance the field of multiscale active materials, leading to innovative energy storage solutions for the future.

**ACKNOWLEDGEMENT**

The author acknowledges the funding support from National Science Foundation (Award Number - 1911900 and 2237990) and Extreme Science and Engineering Discovery Environment (XSEDE) for the computational facilities (Award Number - DMR180013).

**COMPETING INTEREST**

The authors declare no competing financial interest.